\documentclass[nofootinbib]{revtex4}
\usepackage{graphicx}
\usepackage{fancyhdr}
\usepackage{amsmath}
\pdfoutput=1

\newcommand{\vev}[1]{\left\langle#1\right\rangle}
\newcommand{\dis}{\displaystyle}
\newcommand{\al}[1]{\begin{align}#1\end{align}}

\newcommand{\D}{\mathcal{D}}

\newcommand{\U}{\mathcal{U}}

\newcommand{\PRD}[3]{{\it Phys. Rev.} {\bf D{#1}} ({#2}), {#3}}

\newcommand{\PLB}[3]{{\it Phys. Lett. } {\bf B{#1}} ({#2}), {#3}}

\newcommand{\PTP}[3]{{\it Prog. Theor. Phys.} {\bf {#1}} ({#2}), {#3}}

\newcommand{\PTEP}[2]{{\it  Prog. Theor. Exp. Phys.} ({#1}){\bf{ #2}}}

\usepackage{color}

\begin{document}

%Title of paper
\title{CP phase from Higgs's boundary condition}

% Repeat the \author .. \affiliation  etc. as needed
%
% \affiliation command applies to all authors since the last
% \affiliation command. The \affiliation command should follow the
% other information

\author{Y. Fujimoto\footnote{This talk is given by Y.Fujimoto in the conference HPNP2013.}${}^{\dagger}$, K. Nishiwaki${}^{\ddagger}$, M. Sakamoto${}^{\dagger}$}
\affiliation{${}^{\dagger}$Department of Physics, Kobe University, Kobe 657-8501, Japan\\
${}^{\ddagger}$Regional Centre for Accelerator-based Particle Physics, Harich-Chandra Reseach Institute, Allahabad 211 019, India}

\begin{abstract}
We propose a new mechanism to generate a CP phase originating from a non-trivial Higgs vacuum expectation value in an extra dimension. A twisted boundary condition for the Higgs doublet can produce an extra dimensional coordinate-dependent vacuum expectation value containing a CP phase degree of freedom. With this mechanism, we construct a phenomenological model on $S^1$ which can simultaneously and naturally explain the origins of the fermion generations, the quark mass hierarchy and the structure of the Cabibbo-Kobayashi-Maskawa matrix with the CP phase\footnote{This talk is based on Ref.\cite{FujimotoCP}}.
\end{abstract}

%\maketitle must follow title, authors, abstract
\maketitle

\thispagestyle{fancy}

% body of paper here - Use proper section commands
% References should be done using the \cite, \ref, and \label commands
% Put \label in argument of \section for cross-referencing
%\section{\label{}}

%%%%%%%%%%%%%%%%%%%%%%%%%%%%%%%%%%
\section{Introduction}
A quest for the origins of the quark mass hierarchy, the structure of the flavor mixing, and the three generations of the fermions is one of the important tasks in particle physics. Lots of experiments have succeeded in measuring values of the quark masses and the elements of the Cabbibo-Kobayashi-Maskawa (CKM) matrix with good precision. A complex phase in the CKM matrix due to the three generations has been proposed to explain an origin of the CP violation \cite{KM1973}, and the existence of the CP phase has been well established by B physics experiments. However, the Standard Model (SM) does not initiate us into the origins of quark mass hierarchy, the structure of the flavor mixing, and three generations of the fermions even though the SM contains these structures. Thus we consider that there is a more fundamental theory beyond the SM.

In the context of higher dimensional field theories, which are one of the candidates of beyond the SM, we propose a new mechanism to produce a CP phase and construct a five-dimensional (5D)  phenomenological model with $S^1$ compactification which can naturally explain all the flavor structure of the SM, i.e. the origins of the fermion generations, the quark mass hierarchy and the structure of the CKM matrix with the CP phase.
%In our model, the flavor structure of the SM are determined through the geometry of the extra dimension. 
We put point interactions on $S^1$, which are additional boundary points on $S^1$,  to realized the three generations from a single 5D fermion. 
It should be emphasized that 5D Yukawa couplings cannot be the origin of the CP phase in our model because the model contains only a single generation fermions for each 5D quark. A twisted boundary condition (BC) for the Higgs doublet leads to a CP phase degree of freedom. The Higgs VEV with an extra dimensional coordinate dependent phase is a key and will be derived in the next section. We also introduce an extra dimensional coordinate dependent vacuum expectation value (VEV) of a gauge singlet scalar field to realize the quark mass hierarchy. The Robin BC for the gauge singlet scalar field can lead to suitable form of the VEV. The structure of the flavor mixing is determined by the geometry of the extra dimension in our model.

%%%%%%%%%%%%%%%%%%%%%%%%%%%%%%%%%%
\section{Higgs VEV with twisted boundary condition}
First, we discuss the property of the VEV of a $SU(2)_W$ Higgs doublet $H$ with a twisted boundary condition on $S^1$.
The action we consider is
	%%%%%%%%%%%%%%%%%%%%%%%%%%%%%%%%%%
	\begin{align}
	S_H &= \int d^4x \int_0^L dy \Bigl[ {-|\partial_{M} H|^2 + {M}^2 |H|^2 - \frac{\lambda}{2} |H|^4} \Bigr].\label{doubletaction}
	\end{align}
	%%%%%%%%%%%%%%%%%%%%%%%%%%%%%%%%%%
We impose the twisted boundary condition on $H$ as
	%%%%%%%%%%%%%%%%%%%%%%%%%%%%%%%%%%
	\begin{align}
	H(y+L) = e^{i \theta} H(y).\label{twistBC}
	\end{align}
	%%%%%%%%%%%%%%%%%%%%%%%%%%%%%%%%%%
Here, we take the range of $\theta$ as $-\pi < \theta \leq \pi$. We will obtain the VEV of $\langle H(y) \rangle$ minimizing the functional
	%%%%%%%%%%%%%%%%%%%%%%%%%%%%%%%%%%
	\begin{align}
	{\cal E}[H] = \int_0^{L} dy \Bigl[\ |\partial_y H|^2 - {M}^2 |H|^2 + \frac{\lambda}{2} |H|^4 \ \Bigr].\label{functional}
	\end{align}
	%%%%%%%%%%%%%%%%%%%%%%%%%%%%%%%%%%
To find the minimization condition of the functional ${\cal E}$, we introduce ${\cal H}(y)$ as $H(y)=e^{i\frac{\theta}{L}y}{\cal H}(y),\ \ {\cal H}(y+L)={\cal H}(y)$. The VEV of $\langle {\cal H}(y)\rangle$ which minimize the functional ${\cal E}$ will lead us to the VEV of $\langle H(y)\rangle$.  See Ref.\cite{Fujimoto:2012wv} in details. The VEV $\langle H(y) \rangle$ is
given, without any loss of generality, as follows:
\begin{itemize}
\item For $M^2 -\left(\frac{\theta}{L}\right)^2>0$ case 
	%%%%%%%%%%%%%%%%%%%%%%%%%%%%%%%%%%
	\begin{align}	
	\langle H\rangle  =\left\{ \begin{array}{l}
					\frac{v}{\sqrt{2}}e^{i\frac{\theta}{L}y}\left( \begin{array}{c}
													0\\
													1
												      \end{array}\right) \hspace{11em}{\rm for}\ \ -\pi<\theta<\pi\\
					\frac{v}{\sqrt{2}}e^{i\frac{\pi}{L}y}\left( \begin{array}{c}
													0\\
													1
												      \end{array}\right) \ \ {\rm or}\ \ \frac{v}{\sqrt{2}}e^{-i\frac{\pi}{L}y}\left( \begin{array}{c}
													0\\
													1
												      \end{array}\right) \hspace{1em}\ \ \ {\rm for}\ \ \ \theta=\pi
				     \end{array}\right. , \label{doubletVEVform1}
	\end{align}
	%%%%%%%%%%%%%%%%%%%%%%%%%%%%%%%%%%
\item For $M^2 -\left(\frac{\theta}{L}\right)^2<0$ case
	%%%%%%%%%%%%%%%%%%%%%%%%%%%%%%%%%%
	\begin{align}
\vev{H(y)} = \begin{pmatrix} 0 \\ 0 \end{pmatrix}{,} \label{doubletVEVform2}
	\end{align}
	%%%%%%%%%%%%%%%%%%%%%%%%%%%%%%%%%%
\end{itemize}
where $v$ is given by
	%%%%%%%%%%%%%%%%%%%%%%%%%%%%%%%%%%
	\begin{align}	
	\left(\frac{v}{\sqrt{2}}\right)^2 :=|\langle H(y) \rangle |^2 = \frac{1}{\lambda}\left(M^2 -\Bigl(\frac{\theta}{L}\Bigr)^2\right). \label{squared_VEV}
	\end{align}
	%%%%%%%%%%%%%%%%%%%%%%%%%%%%%%%%%%
In the following, we will assume the case of $M^2 -\left(\frac{\theta}{L}\right)^2>0$.
Now we discuss some properties of the derived VEV in Eq.~(\ref{doubletVEVform1}). Differently from the SM, {the VEV possesses $y$-position-dependence} and its broken phase is realized only in the case of {${M}^2 - \left(\frac{\theta}{L}\right)^2 > 0$}{. But} like the SM,
{the squared VEV~(\ref{squared_VEV}) is still constant even though $\langle H(y) \rangle$ depends on $y$}.
This means that after $v/\sqrt{L}$ is set as $246\,\text{GeV}$, where the mass dimension of %the 5D VEV 
$v$ is $3/2$, the same situation {as} the SM %is generated
occurs in the electroweak symmetry breaking (EWSB) sector.
On the other hand, the $y$-dependence of the Higgs VEV
in {Eq.}~(\ref{doubletVEVform1}) is an important consequence for the Yukawa
sector.
Since the VEV of the Higgs doublet appears linearly in each Yukawa term,
the overlap integrals which lead to effective 4D Yukawa couplings 
will produce non-trivial CP phase in the CKM matrix.

We also comment on the Higgs-quarks couplings in our model. 
The profiles of the VEV and the Higgs physical zero mode in our model
are the same as $e^{i \frac{\theta}{L} y}$ up to the coefficients.
This means that the strengths of the couplings are equivalent to those of the SM even though the mode function gets to be $y$-position dependent.
As a result, the decay branching ratios of the Higgs boson are the same as those of the SM.
Possible deviations in the partial widths of the one-loop induced
processes could be small when we take the Kaluza-Klein (KK) scale
around a few TeV.

%%%%%%%%%%%%%%%%%%%%%%%%%%%%%%%%%%
\section{The model with point interactions on $S^1$}
Field localization in extra dimensions is known as an effective way of explaining the quark mass hierarchy and pattern of flavor mixing. For this purpose, we follow the strategy in \cite{Fujimoto:2012wv}, where point interactions are introduced in the bulk space to split and localize fermion profiles and also to produce a $y$-position-dependent VEV with an (almost) exponential shape, which generates the large fermion mass hierarchy.
But in this letter, we set the extra dimension to be a circle $S^1$ not an interval as \cite{Fujimoto:2012wv}. Under the situation, the twisted boundary condition (\ref{twistBC}) is compatible with the geometry.
In the following part, we briefly explain how to construct our model. The 5D action for fermions is given by
	%%%%%%%%%%%%%%%%%%%%%%%%%%%%%%%%%%
	\begin{align}	
	S=\int d^4 x\int^{L}_{0}dy \left[ \bar{Q}(i\Gamma^{M}\partial_{M}+M_{Q})Q
							+\bar{{\cal U}}(i\Gamma^{M}\partial_{M}+M_{{\cal U}}){\cal U}
							+\bar{{\cal D}}(i\Gamma^{M}\partial_{M}+M_{{\cal D}}){\cal D}\right],
	\end{align}
	%%%%%%%%%%%%%%%%%%%%%%%%%%%%%%%%%%
where we introduce  an $SU(2)_W$ doublet $Q$, an up-quark singlet $\U$, 
and a down-type singlet $\D$.
{We note that our model contains only one generation 
for 5D quarks %. However, as we will see soon, 
but each 5D quark produces three generations of the 4D quarks,
as we will see below.}

%When 
We adopt the following BCs for $Q, \U, \D$ %, respectively 
with an infinitesimal {positive} constant %value $\varepsilon$ as
$\varepsilon$ \cite{Fujimoto:2012wv}:
	%%%%%%%%%%%%%%%%%%%%%%%%%%%%%%%%%%
	\begin{align}	
	Q_R &= 0  \hspace{2em} \text{at} \quad y=L_0^{(q)} + \varepsilon,\, L_1^{(q)} \pm \varepsilon,\, L_2^{(q)} \pm \varepsilon,\, L_3^{(q)} - \varepsilon, \label{SMfermion_BC1} \\
{\cal U}_L &= 0  \hspace{2em} \text{at} \quad y=L_0^{(u)} + \varepsilon,\, L_1^{(u)} \pm \varepsilon,\, L_2^{(u)} \pm \varepsilon,\, L_3^{(u)} - \varepsilon, \label{SMfermion_BC2} \\
{\cal D}_L &= 0  \hspace{2em} \text{at} \quad y=L_0^{(d)} + \varepsilon,\, L_1^{(d)} \pm \varepsilon,\, L_2^{(d)} \pm \varepsilon,\, L_3^{(d)} - \varepsilon, \label{SMfermion_BC3}
	\end{align}
	%%%%%%%%%%%%%%%%%%%%%%%%%%%%%%%%%%
	where $\Psi_{R}$ and $\Psi_{L}$ denote the eigenstates of $\gamma^{5}$, i.e.
$\Psi_{R} \equiv \frac{1+\gamma^{5}}{2}\Psi$ and
$\Psi_{L} \equiv \frac{1-\gamma^{5}}{2}\Psi$.
A crucial consequence of the above BCs is that there appear 
three-fold degenerated left- {(right-)handed} zero modes in the 
mode expansions of $Q$ (${\cal U},{\cal D}$) and that they form the three
generations of the quarks.
The details have been given in Ref.~\cite{Fujimoto:2012wv}.
We will not repeat the discussions here.

The fields $Q, \mathcal{U}, \mathcal{D}$  with the BCs in Eqs~(\ref{SMfermion_BC1})--(\ref{SMfermion_BC3}) are KK-decomposed as follows:
	%%%%%%%%%%%%%%%%%%%%%%%%%%%%%%%%%%
	\begin{align}	
	Q(x,y) = \begin{pmatrix} U(x,y) \\ D(x,y) \end{pmatrix} &=
\begin{pmatrix}
\sum_{i=1}^{3} u^{(0)}_{iL}(x) f_{q^{(0)}_{iL}}(y) \\
\sum_{i=1}^{3} d^{(0)}_{iL}(x) f_{q^{(0)}_{iL}}(y)
\end{pmatrix}
+ (\text{KK modes}), \\
\mathcal{U}(x,y) &=
\sum_{i=1}^{3} u^{(0)}_{iR}(x) f_{u^{(0)}_{iR}}(y) + (\text{KK modes}), \\
\mathcal{D}(x,y) &=
\sum_{i=1}^{3} d^{(0)}_{iR}(x) f_{d^{(0)}_{iR}}(y) + (\text{KK modes}).
	\end{align}
	%%%%%%%%%%%%%%%%%%%%%%%%%%%%%%%%%%	
Here the zero mode functions are obtained in the following forms:
	%%%%%%%%%%%%%%%%%%%%%%%%%%%%%%%%%%
	\begin{align}
	f_{q^{(0)}_{iL}}(y) &= 
		\mathcal{N}_{i}^{(q)} e^{M_{Q} (y-L_{i-1}^{(q)})} \Big[ \theta(y-L_{i-1}^{(q)}) \theta(L_{i}^{(q)}-y) \Big]
		&& \text{in\ }[L_0^{(q)},L_3^{(q)}], \label{doubletwavefunction} \\
	f_{u^{(0)}_{iR}}(y) &= 
		\mathcal{N}_{i}^{(u)} e^{-M_{\mathcal{U}} (y-L_{i-1}^{(u)})} \Big[ \theta(y-L_{i-1}^{(u)}) \theta(L_{i}^{(u)}-y) \Big]
		&& \text{in\ }[L_0^{(u)},L_3^{(u)}], \label{upsingletwavefunction} \\
	f_{d^{(0)}_{iR}}(y) &= 
		\mathcal{N}_{i}^{(d)} e^{-M_{\mathcal{D}} (y-L_{i-1}^{(d)})} \Big[ \theta(y-L_{i-1}^{(d)}) \theta(L_{i}^{(d)}-y) \Big]
		&& \text{in\ }[L_0^{(d)},L_3^{(d)}], \label{downsingletwavefunction}
	\end{align}
	%%%%%%%%%%%%%%%%%%%%%%%%%%%%%%%%%%
where 
	%%%%%%%%%%%%%%%%%%%%%%%%%%%%%%%%%%
	\begin{align}
	\Delta L^{(l)}_{i} &= L^{(l)}_{i} - L^{(l)}_{i-1} \qquad
(\text{for } i=1,2,3;\ l=q,u,d),
	\end{align}
	%%%%%%%%%%%%%%%%%%%%%%%%%%%%%%%%%%
\vspace{-7mm}
%
	%%%%%%%%%%%%%%%%%%%%%%%%%%%%%%%%%%
	\begin{align}
	\mathcal{N}_{i}^{(q)} = \sqrt{\frac{2M_{Q}}{e^{2M_{Q} \Delta L_{i}^{(q)} }-1}}, \quad
	\mathcal{N}_{i}^{(u)} = \sqrt{\frac{2M_{\mathcal{U}}}{1-e^{-2M_{\mathcal{U}} \Delta L_{i}^{(u)} }}}, \quad
	\mathcal{N}_{i}^{(d)} = \sqrt{\frac{2M_{\mathcal{D}}}{1-e^{-2M_{\mathcal{D}} \Delta L_{i}^{(d)} }}}.
\label{wavefunctionconventions}
	\end{align}
	%%%%%%%%%%%%%%%%%%%%%%%%%%%%%%%%%%
$\mathcal{N}_{i}^{(q)}, \mathcal{N}_{i}^{(u)}, \mathcal{N}_{i}^{(d)}$ are the wavefunction normalization factors for $f_{q^{(0)}_{iL}}, f_{u^{(0)}_{iL}}, f_{d^{(0)}_{iL}}$, respectively.

Since the length of the total system is universal, 
$L^{(l)}_{3} - L^{(l)}_{0}\ (l=q,u,d)$ should be equal to the
circumference of $S^{1}$, i.e.
	%%%%%%%%%%%%%%%%%%%%%%%%%%%%%%%%%%
	\begin{align}
	L := L^{(q)}_{3} - L^{(q)}_{0} = L^{(u)}_{3} - L^{(u)}_{0} = L^{(d)}_{3} - L^{(d)}_{0}.
	\label{totallength}
	\end{align}
	%%%%%%%%%%%%%%%%%%%%%%%%%%%%%%%%%%
Note that all the mode functions in Eqs.~(\ref{doubletwavefunction})--(\ref{downsingletwavefunction}) (and a form of a singlet VEV in Eq.~(\ref{exactHiggsVEVform})) are periodic with the common period $L$, whereas we do not indicate that thing explicitly in Eqs.~(\ref{doubletwavefunction})--(\ref{downsingletwavefunction}).

	%%%%%%%%%%%%%%%%%%%%%%%%%%%%%%%%%
	\begin{figure}[t]
	\centering
	%\vspace{-22mm}
	\includegraphics[width=0.7\columnwidth]{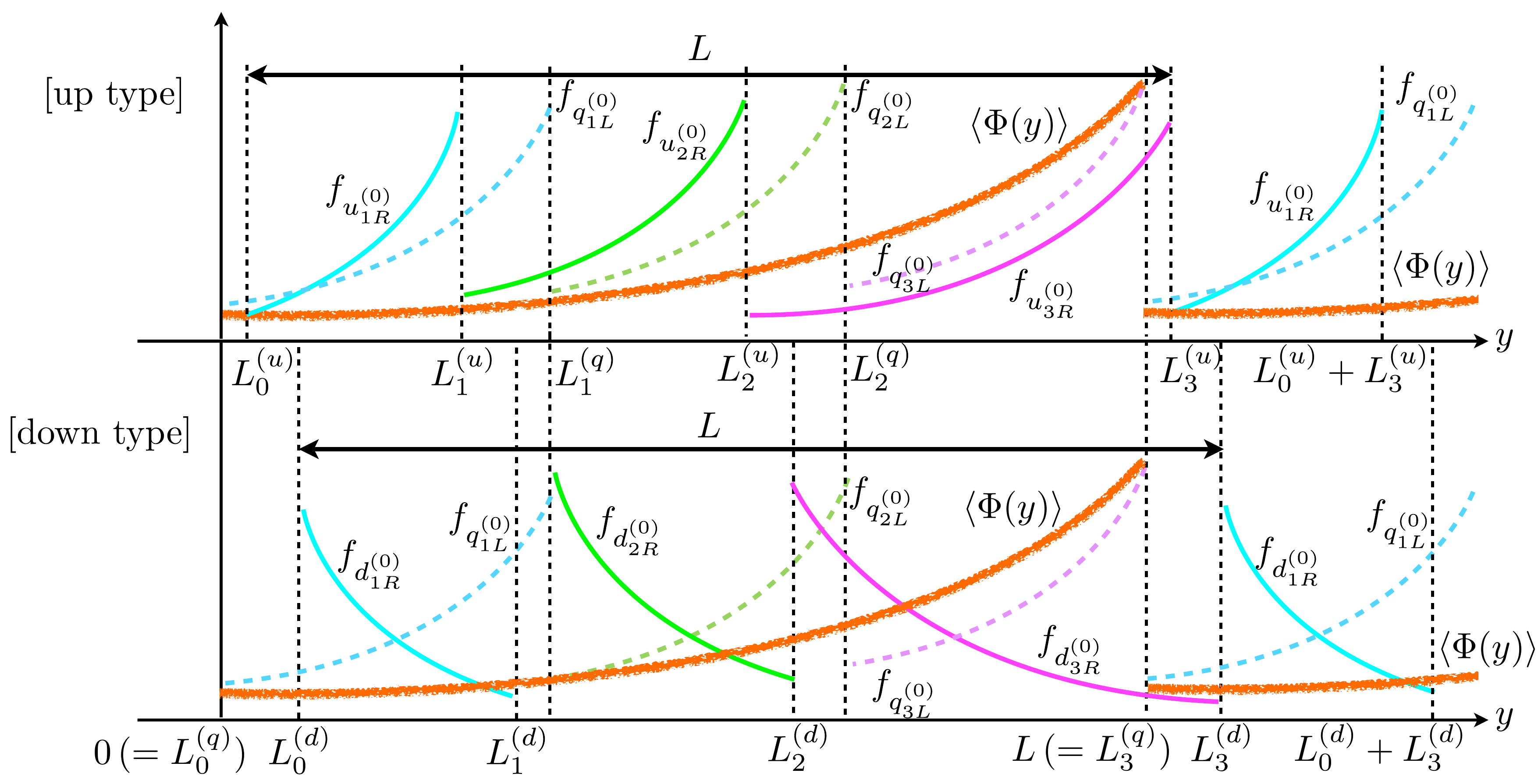}
	%\vspace{-9mm}
	\caption{{\footnotesize 
	%An outlines of the wavefunctions profiles.
	The wavefunction profiles of the quarks and the VEV of $\Phi(y)$ are
	schematically depicted. 
	Here we take $L_0^{(q)}=L_0^{(\Phi)}=0$.
	%set $L_0^{(q)}$ and $L_0^{(\Phi)}$ as zero.
	Note that all the profiles %has
	have the periodicity along $y$ with the same period $L$.
	Differently from the model on an interval in Ref.~\cite{Fujimoto:2012wv}, we can find the $(1,3)$ elements of the mass matrices due to the 	periodicity along $y$-direction.}
	}
	\label{totalprofile_pdf}
	\end{figure}
	%%%%%%%%%%%%%%%%%%%%%%%%%%%%%%%%%

In this model, the large mass hierarchy is naturally explained with the following Yukawa sector
\begin{align}
S_{{\cal Y}} = \int d^4 x \int_{0}^{{L}} dy \bigg\{ \Phi \Big[
		- {\cal Y}^{(u)} \overline{Q} (i \sigma_2 H^{\ast}) {\cal U} - {\cal Y}^{(d)} \overline{Q} H {\cal D}  \Big]
		+ \text{h.c.}
		\bigg\},
\label{higherdimensionalYukawa}
\end{align}
where ${\cal Y}^{(u)}/{\cal Y}^{(d)}$ is the Yukawa coupling for up/down type quark;
%$Q$, $\U$ and $\D$ are $SU(2)_W$ doublet, up-type and down-type singlets, respectively;
$H$ and $\Phi$ are an $SU(2)_W$ scalar doublet and a singlet.
%$\Y^{(u)}$ and $\Y^{(d)}$ can be complex values, {but they cannot act as physical CP phase d.o.f.}
It should be noted that 
although the Yukawa couplings ${\cal Y}^{(u)}$ and ${\cal Y}^{(d)}$ can be complex, 
they cannot be an origin of the CP phase of the CKM matrix
because our model contains only a single quark generation, so that
the number of the 5D Yukawa couplings is not enough
to produce a CP phase in the CKM matrix.
%because the complex phases of $\Y^{(u)}$ and $\Y^{(d)}$ are not
%physical and can be absorbed into the phases of the quark fields.
An schematic figure of our system is {depicted} in Fig.~\ref{totalprofile_pdf}.
Note that the five terms of $\overline{Q} (i \sigma_2 H^{\ast}) {\cal U}, \overline{Q} H {\cal D}, \Phi \overline{Q} Q, \Phi \overline{{\cal U}} {\cal U}, \Phi \overline{{\cal D}} {\cal D}$ with the Pauli matrix $\sigma_2$ are %precluded 
excluded by introducing a discrete symmetry 
$H \rightarrow -H, \Phi \rightarrow -\Phi$.
$\Phi$ is a gauge singlet and there is no problem with gauge universality violation.

The 5D action and the BCs for $\Phi$ are %given
assumed to be of the form \cite{Fujimoto:2012wv, Fujimoto:2011kf}
\begin{align}
S_{\Phi} = \int d^4 x \int_{0}^{L} dy \biggl\{ \Phi^{\dagger} \Bigl( \partial_M \partial^M - M_{\Phi}^2 \Bigr)\Phi
		- \frac{\lambda_{\Phi}}{2} \Bigl(\Phi^\dagger \Phi\Bigr)^2 \biggr\}, 
\label{Phi_action}
\end{align}
\vspace{-5mm}
\begin{align}
\Phi + L_+ \partial_y \Phi &= 0 \hspace{4em} \text{at} \ \ \ \  y=L_{0}^{(\Phi)} + \varepsilon, \notag \\
\Phi - L_- \partial_y \Phi &= 0 \hspace{4em} \text{at}  \ \ \ \  y=L_3^{(\Phi)} - \varepsilon,
\label{RobinBC}
\end{align}
where $L_{\pm}$ can take values in the range of $-\infty \leq L_{\pm} \leq \infty$ and $L_{0}^{(\Phi)}$ and $L_{3}^{(\Phi)}$ indicate the locations of the two ``end points" of the singlet.
The VEV of $\Phi$ with the BCs, named Robin BCs, in Eq.~(\ref{RobinBC}) is expressed in terms of Jacobi's elliptic functions in general and its phase structure {has been discussed} in Ref~\cite{Fujimoto:2011kf}.
We adopt a specific form in the region $[L_{0}^{(\Phi)} + \varepsilon, L_{3}^{(\Phi)} - \varepsilon]$\cite{Fujimoto:2012wv}:

\begin{align}
\langle \Phi(y)\rangle =\left [ \frac{M_{\Phi}}{\sqrt{\lambda_{\Phi}}} \Bigl\{ \sqrt{1 + X}-1\Bigr\}^{\frac{1}{2}} \right]\times \frac{1}{{\rm cn} \left( M_{\Phi}\{1+X\}^{1/4} (y-y_{0}), \sqrt{\frac{1}{2}(1+\frac{1}{\sqrt{1+X}})}\ \right)}, \hspace{1em}\Bigl(X:= \frac{4\lambda_\Phi |Q|}{M^4_\Phi}\Bigr).
\label{exactHiggsVEVform}
\end{align}

Here $y_0$ and $Q$ are parameters which appear after integration on $y$ and we focus on the choice of $Q<0$.
We note that the values of $y_0$ and $Q$ are automatically determined after choosing those of $L_{\pm}$.
As shown in Ref.~\cite{Fujimoto:2012wv}, 
we get the form of $\langle \Phi(y)\rangle$ to be an (almost) exponential function
of $y$ by
choosing suitable parameter configurations.
Although there is a discontinuity in the wavefunction profile of $\langle \Phi \rangle$ between
$y=L_{0}^{(\Phi)}+\varepsilon$ and $y=L_{3}^{(\Phi)}-\varepsilon$ in Eqs.~(\ref{RobinBC}),
this type of BCs is derived from the variational principle on $S^1$
and leads to no inconsistency~\cite{Fujimoto:2011kf}.
The BCs for the 5D $SU(3)_C, SU(2)_W, U(1)_Y$ gauge bosons $G_M, W_M, B_M$ are selected as
\begin{align}
G_{M}|_{y=0} = G_{M}|_{y=L},\ \ \ \ 
\partial_y G_{M}|_{y=0} = \partial_y G_{M}|_{y=L},
\label{gluon_BC}
\end{align}
where we only show the $G_M$'s case.
In this configuration, we %can find candidates for 
obtain the SM gauge bosons in zero modes.
Based on the discussion in Section~II, we conclude that the W and Z bosons become massive and {their masses} are suitably created through ``our" Higgs mechanism as $m_W \simeq 81\,\text{GeV}, m_Z \simeq 90\,\text{GeV}$.
We mention that, on $S^1$ geometry, $G^{(0)}_y$, $W^{(0)}_y$, and $B^{(0)}_y$ %exist as physical massless modes 
would exist as massless 4D scalars at the tree level, 
but {they will become massive via quantum corrections and are
expected to be uplifted to near KK states.} {We will discuss those modes
in another paper.}
We should note that in our model on $S^1$ with point interactions, the 5D gauge symmetries are intact under the BCs (\ref{twistBC}),(\ref{SMfermion_BC1})-(\ref{SMfermion_BC3}),(\ref{RobinBC}),(\ref{gluon_BC}).

\section{Results}
In this section, we %strive 
would like to find a set of parameter {configurations in} which the quark mass hierarchy and the structure of the CKM matrix with the CP phase are derived naturally.
In %our 
the following analysis, we %scale 
rescale all the %massive 
dimensional valuables by the $S^1$ circumference $L$ 
to make them dimensionless and the %scaled
rescaled %values 
valuables are indicated with the tilde $\tilde{{}}$\,.

We set the parameters concerning the scalar singlet $\Phi$ as
\begin{align}
{\tilde{M}_\Phi} = 8.67, \quad \tilde{y}_0 = - 0.1, \quad {\tilde{\lambda}_\Phi} = 0.001, \quad |\tilde{Q}| = 0.001,
\label{asetofHiggsparameters}
\end{align}
where the VEV profile becomes an (almost) exponential
function of $y$, 
which is suitable for generating the large mass hierarchy. In this case, the values of $L_{\pm}$ %describing the singlet's BC 
in Eq.~(\ref{RobinBC}) correspond to
\begin{align}
\frac{1}{\tilde{L}_+} = - 6.07, \quad \frac{1}{\tilde{L}_-} = 8.69,
\label{Lplusminusvalues}
\end{align}
where the broken phase is realized %in the singlet~
\cite{Fujimoto:2012wv}.{}

As in the previous analysis %in~
\cite{Fujimoto:2012wv}, the signs of the fermion bulk masses are assigned as
$M_Q > 0, M_{{\cal U}} < 0, M_{{\cal D}} > 0$ to make much larger overlapping in up quark sector than in down ones for top mass.
Here we assume the positions of the two ``end points" of both the quark doublet and the scalar singlet are the same
\begin{align}
L_0^{(q)} = L_0^{(\Phi)} = 0,\quad
L_3^{(q)} = L_3^{(\Phi)} = L,
\label{L_assumption}
\end{align}
where we set $L_0^{(q)}$ and $L_0^{(\Phi)}$ as zero.
In addition, we also assume that the orders of the positions of point interactions are settled as
\begin{align}
0 < L_0^{(u)} < L_1^{(u)} < L_1^{(q)} < L_2^{(u)} < L_2^{(q)} < L < L_3^{(u)}, \notag \\
0 < L_0^{(d)} < L_1^{(d)} < L_1^{(q)} < L_2^{(d)} < L_2^{(q)} < L < L_3^{(d)}.
\label{positionorder1}
\end{align}

Here our up quark mass matrix ${\mathcal{M}}^{(u)}$ and that of down ones ${\mathcal{M}}^{(d)}$ take the forms
\begin{align}
\mathcal{M}^{(u)} =
\begin{bmatrix}
m^{(u)}_{11} & m^{(u)}_{12} & m^{(u)}_{13} \\
0 & m^{(u)}_{22} & m^{(u)}_{21} \\
0 & 0 & m^{(u)}_{33}  
\end{bmatrix}, \ \ \ \ 
\mathcal{M}^{(d)} =
\begin{bmatrix}
m^{(d)}_{11} & m^{(d)}_{12} & m^{(d)}_{13} \\
0 & m^{(d)}_{22} & m^{(d)}_{21} \\
0 & 0 & m^{(d)}_{33}  
\end{bmatrix},
\label{massmatrix1}
\end{align}
where the row (column) index of the mass matrices {shows the generations of the left- (right-)handed fermions}, respectively.
Differently from the model on an interval in Ref.~\cite{Fujimoto:2012wv}, {the $(1,3)$ elements of the mass matrices are allowed geometrically} due to the periodicity along $y$-direction.

The parameters which we use for calculation are
\begin{align}
\begin{array}{llll}
\tilde{L}_{0}^{(q)} = 0, & \tilde{L}_{1}^{(q)} = 0.298, & \tilde{L}_{2}^{(q)} = 0.659, &
		\tilde{L}_{3}^{(q)} = 1,  \\
\tilde{L}_{0}^{(u)} = 0.0245, & \tilde{L}_{1}^{(u)} = 0.0260, & \tilde{L}_{2}^{(u)} = 0.520, &
		\tilde{L}_{3}^{(u)} = {1.03},  \\
\tilde{L}_{0}^{(d)} = 0.0703, &
		\tilde{L}_{1}^{(d)} = 0.178, & \tilde{L}_{2}^{(d)} = 0.646, &
		\tilde{L}_{3}^{(d)} = {1.07}, \\
\tilde{M}_Q = 0.654, & \tilde{M}_{{\cal U}} = - 0.690, & \tilde{M}_{{\cal D}} = 0.595, &
		\theta = 3.0,
\end{array}
\label{a_setofsolution}
\end{align}
where %we note that 
the twist angle $\theta$ is a {dimensionless} value and should be within the range {$-\pi < \theta \leq \pi$}.
We should remind that in our system, the EWSB is only realized on the condition of {${M}^2 - \left(\frac{\theta}{L}\right)^2 > 0$} as in Eq.~(\ref{doubletVEVform1}). %and (\ref{doubletVEVform2}).
Recently, the ATLAS and CMS experiments have announced that the physical Higgs mass is around $126\,\text{GeV}$ with $5\sigma$ confidence level~\cite{2012gk, 2012gu}.
{$\tilde{\lambda}$} is $0.262$ irrespective of the value of $L$, {while {$\tilde{M}$} is slightly dependent on the value of $L$ as $3.01303$ $(3.00052)$ in the case of $M_{\text{KK}}=2\,\text{TeV}$ ($M_{\text{KK}}=10\,\text{TeV}$), where $M_{\text{KK}}$ is a typical scale of the KK mode and defined as $2\pi/L$}. 
Here %somewhat of 
some {tuning} is required to obtain the suitable values realizing the EWSB.

After the %diagonalizing 
diagonalization of the two mass %matrix
matrices, the quark masses are evaluated as
\begin{equation}
\begin{array}{lll}
m_{\text{up}} = 2.06\,\text{MeV}, &
	m_{\text{charm}} = 1.25\,\text{GeV}, &
	m_{\text{top}} = 174\,\text{GeV},  \\[0.1cm]
m_{\text{down}} = 4.91\,\text{MeV}, &
	m_{\text{strange}} = 102\,\text{MeV}, &
	m_{\text{bottom}} = 4.18\,\text{GeV},  \\
\\
\dis \frac{m_{\text{up}}}{m_{\text{up}}|_{\text{exp.}}} = 0.897, &
	\dis \frac{m_{\text{charm}}}{m_{\text{charm}}|_{\text{exp.}}} = 0.978, &
	\dis \frac{m_{\text{top}}}{m_{\text{top}}|_{\text{exp.}}} = 1.00, \\[0.4cm]
\dis \frac{m_{\text{down}}}{m_{\text{down}}|_{\text{exp.}}} = 1.02, &
	\dis \frac{m_{\text{strange}}}{m_{\text{strange}}|_{\text{exp.}}} = 1.07, &
	\dis \frac{m_{\text{bottom}}}{m_{\text{bottom}}|_{\text{exp.}}} = 1.00,
\end{array}
\label{obtainedquarkmass}
\end{equation}

and the absolute values of the {CKM matrix elements} are given as
\al{
|V_{\text{CKM}}|=
\begin{bmatrix}
		0.971 & 0.238 &  0.00318 \\
		0.238 & 0.970 & 0.0372 \\
		0.00829 & 0.0364 & 0.999
\end{bmatrix},\quad
\left| \frac{V_{\text{CKM}}}{V_{\text{CKM}}|_{\text{exp.}}} \right|=
\begin{bmatrix}
		0.997 & 1.06 &  0.906 \\
		1.06 & 0.997 & 0.902 \\
		0.957 & 0.900 & 1.00
\end{bmatrix}.
\label{obtainedCKMmatrix}
}
The Jarlskog parameter $J$ is
\al{
J = 2.56 \times 10^{-5},\quad \frac{J}{J|_{\text{exp.}}} = 0.865,
}
where we also provide the differences from the latest experimental values in Ref.~\cite{Beringer:1900zz}.
All the deviations from the latest experimental values are within about $15\%$ and we can conclude that the situation of the SM is suitably generated.

\section{Summary and discussion}
In this letter, {we %propose 
proposed a new mechanism for generating CP phase via 
the Higgs vacuum expectation value originating from geometry 
of an extra dimension.
%Allowing a 
A twisted boundary condition for the Higgs doublet %drives us into 
has been found to lead to %an extra-dimension-coordinate-dependent 
an extra dimensional coordinate-dependent VEV 
%containing 
with a non-trivial CP phase degree of freedom.
This mechanism is useful for realizing CP violation in an extra-dimensional model. % without a mechanism to generate CP phase. 

As an application of this idea, we %construct 
have constructed a phenomenological model with an extra dimension which can simultaneously and naturally explain the origin of the fermion generations, 
the quark mass hierarchy, and the CKM structure with the CP phase based on~\cite{Fujimoto:2012wv}.
%Introducing and shifting the positions of the 
The point interactions %of the 5D quarks 
realize the {three fermion generations} and the situation where all the quark profiles are split and localized.}
With the help of the almost exponential 
function of the scalar VEV, %profile of the scalar singlet, 
which appears in the Yukawa sector, we can generate 
%the phenomenologically-preferred 
the phenomenologically-desirable circumstances 
where {all the flavor structures are realized with good precision and almost all dimensionless scaled parameters take values of natural $\mathcal{O}(10)$ magnitudes.}

One of the most important remaining {tasks} is to construct a model which can explain both of the quark and lepton flavor structures simultaneously.
%Here, global analysis of the parameter space is helpful to {grow our understanding of the system} with scores of point interactions.
Then, it is necessary to explain why the neutrino masses are so light
and the flavor mixings in the lepton sector are large.
The result will be reported elsewhere.

Another {important} topics is the stability of the system.
Our system {is} possibly threatened with instability. %and some 
Some {mechanisms} will be required to stabilize the moduli representing the positions of point interactions (branes).
%{In a simply-connected space of $S^1$, the other presentiment of jeopardizing the system is coming from Hosotani mechamism~\cite{Hosotani:1983xw,Hosotani:1988bm}.
%The zero modes of the $y$-components of 5D gluon and photon are physical and massless in our construction. Therefore, if they obtain nonzero VEVs at the quantum level, the $SU(3)_C$ and/or $U(1)_{EM}$ gauge symmetries are broken.
%One way to circumvent this possibility is to introduce many additional 5D matters without zero mode. We will cultivate this issue concretely in near future.}
{In a multiply-connected space of $S^{1}$, there is another
origin of gauge symmetry breaking i.e. the Hosotani mechanism.
Since further gauge symmetry breaking causes a problem
in our model, we need to insure that the Hosotani mechanism
does not occur. To this end, we might introduce additional
5D matter to prevent zero modes of $y$-components of gauge
fields from acquiring non-vanishing VEVs. We will leave
{those issues} in future work.}

\begin{acknowledgments}
We would like to thank HPNP2013 organizers for giving Y.F. the opportunity of presentation. The authors would also like to thank T.Kugo for valuable discussions. K.N. is partially supported by funding available from the Department of Atomic Energy, Government of India for the Re- gional Centre for Accelerator-based Particle Physics (RECAPP), Harish-Chandra Research Insti- tute. This work is supported in part by a Grant-in-Aid for Scientific Research (No. 22540281 and No. 20540274 (M.S.)) from the Japanese Ministry of Education, Science, Sports and Culture.
\end{acknowledgments}

\bigskip % extra skip inserted
% Create the reference section using BibTeX:
%\bibliography{basename of .bib file}

\end{document}